\documentclass[conference]{IEEEtran}
\IEEEoverridecommandlockouts
\usepackage{bm}
\usepackage{multirow} 
\usepackage{multicol} 
\usepackage{booktabs}
\usepackage{cite}
\usepackage{amsmath,amssymb,amsfonts}
\usepackage{algorithmic}
\usepackage{graphicx}
\usepackage{textcomp}
\usepackage{xcolor}
\def\BibTeX{{\rm B\kern-.05em{\sc i\kern-.025em b}\kern-.08em
    T\kern-.1667em\lower.7ex\hbox{E}\kern-.125emX}}
\begin{document}

\title{Graph Neural Network and Superpixel Based Brain Tissue Segmentation (Corrected Version)*\thanks{*The original version of this paper was accepted and presented at 2022 International Joint Conference on Neural Networks (IJCNN) \cite{IJCNNGNNSEG}. This version corrects the mistakes in Figs. \ref{fig53} and \ref{fig54}.}}

\author{\IEEEauthorblockN{Chong Wu \& Zhenan Feng \& Houwang Zhang \& Hong Yan}
\IEEEauthorblockA{\textit{Department of Electrical Engineering and Centre for Intelligent Multidimensional Data Analysis}\\
\textit{City University of Hong Kong}\\
Kowloon, Hong Kong SAR \\
chongwu2-c@my.cityu.edu.hk \& zhenafeng2-c@my.cityu.edu.hk \& houwang.zhang@my.cityu.edu.hk \& h.yan@cityu.edu.hk}
}

\maketitle

\begin{abstract}
Convolutional neural networks (CNNs) are usually used as a backbone to design methods in biomedical image segmentation. However, the limitation of receptive field and large number of parameters limit the performance of these methods. In this paper, we propose a graph neural network (GNN) based method named GNN-SEG for the segmentation of brain tissues. Different to conventional CNN based methods, GNN-SEG takes superpixels as basic processing units and uses GNNs to learn the structure of brain tissues. Besides, inspired by the interaction mechanism in biological vision systems, we propose two kinds of interaction modules for feature enhancement and integration. In the experiments, we compared GNN-SEG with state-of-the-art CNN based methods on four datasets of brain magnetic resonance images. The experimental results show the superiority of GNN-SEG.
\end{abstract}


\section{Introduction}
\label{sec:intro}
Convolutional neural networks (CNNs) have gain great breakthroughs in the field of computer vision and image processing \cite{minaee2021image} recently. Especially, since the fully convolutional neural network (FCNN) was proposed for semantic segmentation, a variety of variants have been designed for segmentation tasks \cite{long2015fully}. Among them, Unet and its variants are most widely used in the field of medical image segmentation due to its excellent performance \cite{ronneberger2015u}. 

However, the success of above CNN based methods is limited by the size of receptive field, which results in global information loss \cite{li2020spatial}. A potential way to solve above problem is to design larger and deeper networks. But this will result in huge parameters. And long-range context relationships in the image also cannot always be captured by the stacking local cues precisely \cite{li2020spatial}, which is particularly fatal for some pixel-level tasks, such as semantic segmentation \cite{chen2017deeplab}.

Recently, with the development of graph representation learning, graph neural network (GNN) becomes a powerful technique for modeling pairwise relations \cite{xu2019spatial}. For a medical image, there exists many tree-like structures such as the distribution of gray matter and white matter in brain magnetic resonance images (MRIs). How to segment these tissues is a challenging task, because branches of different tissues are intertwined and vary in size and orientation. Some variants of Unet (CNN based methods) have been proposed to segment brain tissues \cite{sun20193d,zhang2021deep}. However, these methods have some defects like missing various small terminal branches and having huge parameters. Compared with these CNN based methods, GNN has advantages in learning structural information of the image with less parameters. Hence, in this paper, we propose a GNN based method named GNN-SEG for brain tissue segmentation. GNN-SEG consists of two parts: (1) structural relation learning model; (2) pixel-level classifier. It can construct relationships with an awareness of spatial information belong to different tissues and efficiently learn the graph structure of brain from the training slices of brain MRIs. Inspired by the interaction mechanism in biological vision systems, we propose two interaction modules to enhance discrimination ability of features. The contributions of this paper can be concluded into three parts:
\begin{itemize}
\item GNN is introduced in medical image segmentation to tackle the problem of limitation of receptive field in CNN based methods.

\item Inspired by the function of interaction in biological vision systems, two kinds of interaction modules have been proposed to enrich the information of features and enhance the discrimination ability.

\item Compared with mainstream CNN based methods, a more accurate method which can extract rich structural information for the segmentation of brain tissues has been proposed.
\end{itemize}

\section{Methods}
\begin{figure*}[htbp]
	\centering
	\includegraphics[width=\linewidth]{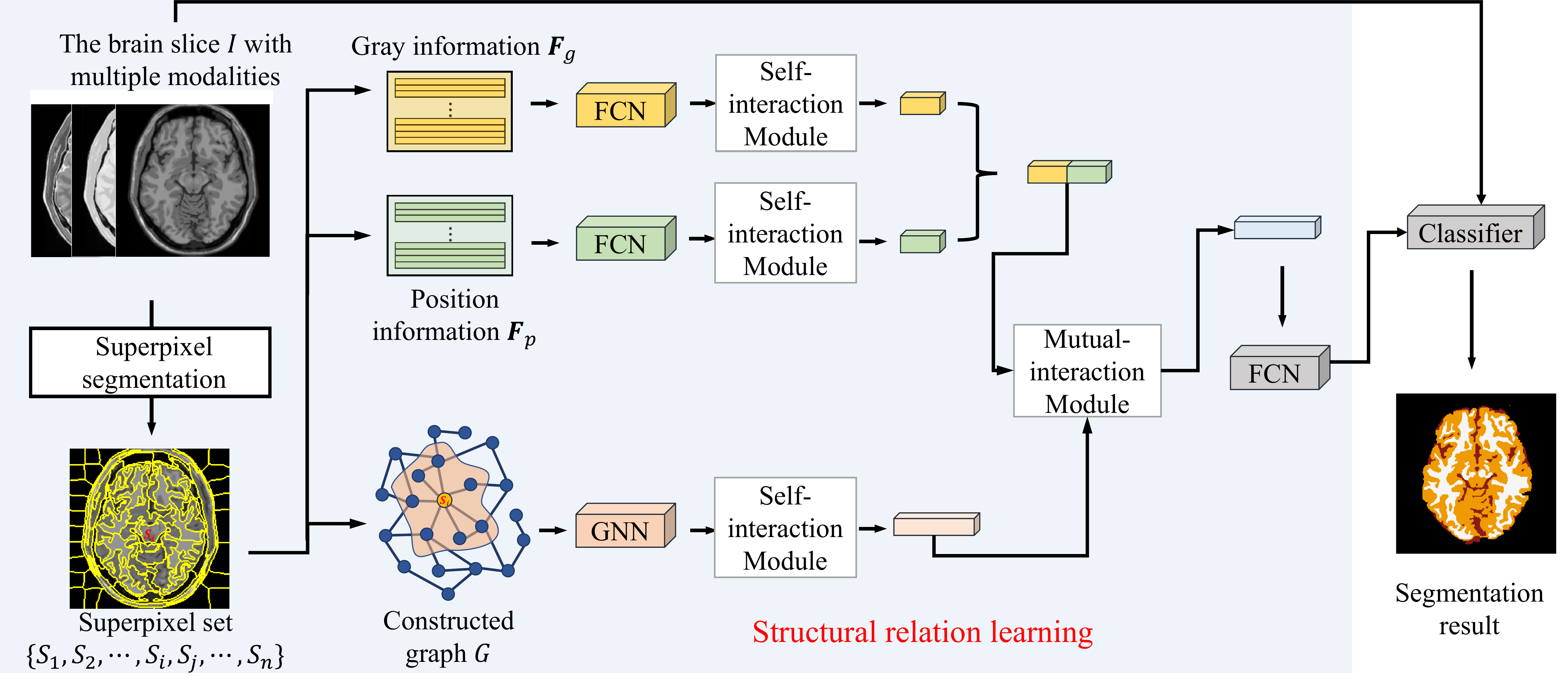}
	\caption{An overview of GNN-SEG. Green and yellow FCN (fully connected network) boxes consist of 2 fully connected layers (hidden unit sizes for 1st-2nd layers are: 20, 100, respectively). Both 2 self-interaction modules after green and yellow FCN boxes consist of 1 fully connected layer with a hidden unit size of 100. GNN denotes a graph neural network (the default GNN in GNN-SEG is graph attention network \cite{velivckovic2017graph}). It has 2 layers (hidden unit sizes for 1st-2nd layers are: 500, 10, respectively) and its multi-head size is 5. Self-interaction module after GNN consists of 1 fully connected layer with a hidden unit size of 10. Mutual-interaction module consists of 2 independent fully connected layers (hidden unit sizes are 200, 10 respectively). Grey box FCN consists of 4 fully connected layers (1st-4th hidden unit sizes: 100, 100, 50, 1). The classifier used in this paper is a pretrained Attention Unet \cite{oktay2018attention}.}
	\label{fig1}
\end{figure*}

The framework of GNN-SEG is as Fig. \ref{fig1} shows. It consists of two parts: (1) structural relation learning model; (2) pixel-level classifier. For structural relation learning of a given brain slice with multiple modalities, we first segment it into superpixels, and then construct a graph of the obtained superpixel nodes. After that, the constructed graph will be input into a GNN (graph attention network \cite{velivckovic2017graph}) to obtain its graph embeddings. Hence, GNN-SEG can learn structural relationship of the brain slice. Besides, two kinds of feature maps of nodes are fed into two fully connected networks to learn color information and spatial information of each node respectively. Then the embedding vectors of nodes and graph are enhanced by self interaction module and mutual interaction module. Next the enhanced vectors are concatenated together. Finally, the concatenated vectors will be processed by a fully connected network to obtain the new structural relation features for each superpixel and as the additional channel information for a pixel-level classifier (Attention Unet \cite{oktay2018attention}) to do pixel-level classification. Next, we will introduce the details of each part of GNN-SEG.

\subsection{Graph Construction}
There are two advantages for segmenting an image into superpixels. First, compared with pixel-level processing, taking superpixel as a basic unit can greatly reduce computational cost. Second, superpixel, which refers to sub regions grouped with some directly connected pixels in an image, can maintain local structural information of the image \cite{fslic}. Therefore, in contrast to pixel, superpixel can preserve more information of images \cite{fslic}. And features of an image like shape, color, and texture can be directly obtained in superpixel. For a given brain slice $I$, it consists of multiple modalities, and among them, some modalities are clearer and have better discrimination for different brain tissues, hence we select the clearest modality for superpixel segmentation. As to segmentation method, we select SNIC \cite{achanta2017superpixels} because of its good segmentation accuracy and running speed. An ablation study of the effect of different superpixel methods on the performance of GNN-SEG is shown in the Section \ref{expp}.

After obtaining superpixel segmentation, we need to construct a graph $G$ based on these superpixel nodes of $I$ for structural relation learning. The principle for construction is using the neighboring relationship of these nodes. For a superpixel $S_i$, it has edges $e_{i,j} \in \mathbb{E}, j\in[1,n], j \neq i$ with neighboring superpixel $S_j$. The weight for each edge is set to $1$, and $n$ is the total number of superpixels. All superpixels form the node set $\mathbb{S} = \{S_1,S_2, ..., S_n\}$. Then we can get the graph $G = (\mathbb{S}, \mathbb{E})$.

Besides, for each superpixel node $S_i$, we extract its mean of gray values (gray value in each modality of a slice) and position values ($x$ and $y$) of its pixels to construct the feature matrices $\bm{F}_g$ and $\bm{F}_p$, respectively. The illustration for these steps can be viewed in Fig. \ref{fig1}.

\subsection{Graph Neural Network}
We use graph neural networks (GNN) to learn structural relations. In this paper, the GNN need to be generalized to completely unseen graphs. Hence, a GNN which can do inductive learning is needed. Graph attention network (GAT) \cite{velivckovic2017graph} is widely used in inductive graph representation learning due to its high performance. So we select GAT as the default graph neural network and other inductive learning GNNs are also available to replace GAT in this paper. In the Section \ref{expp}, we show an ablation study of the effect of different GNNs on the performance of GNN-SEG. For the graph $G$ of the slice $I$, graph embeddings $\bm{h}^{l+1}$ of the $l+1$-th layer can be obtained using formulas \cite{velivckovic2017graph} as follows,
\begin{equation}
\label{ex1}
\bm{r}^{l}_{[S_j,:]} = \bm{h}_{[S_j,:]}^{l}\textbf{\emph{W}}^{l},\ S_j \in \hat{\mathbb{S}}_{i},
\end{equation}
\begin{equation}
\label{ex2}
\theta_{[S_i,S_j]}^{l+1} = \frac{\exp\left(\sigma\left({\rm cat} \left(\bm{r}_{[S_i,:]}^{l},\bm{r}_{[S_j,:]}^{l} \right) \textbf{\emph{a}}^{l+1} \right)\right)}{\sum_{S_j \in \hat{\mathbb{S}}_{i}}\exp\left(\sigma\left({\rm cat} \left(\bm{r}_{[S_i,:]}^{l},\bm{r}_{[S_j,:]}^{l} \right) \textbf{\emph{a}}^{l+1} \right)\right)},
\end{equation}
\begin{equation}
\label{ex3}
\bm{h}^{l+1}_{[S_i,:]} = \sigma (\sum_{S_j \in \hat{\mathbb{S}}_{i}}\theta_{[S_i,S_j]}^{l+1}\bm{r}^{l}_{[S_j,:]}),
\end{equation}
where, $\bm{h}^{l+1}_{[S_i,:]}$ and $\bm{h}^{l}_{[S_i,:]}$ are the high-level feature vectors of the superpixel node $S_i$ obtained by edge attention in the $l+1$-th layer and $l$-th layer respectively, ${\rm cat}()$ is a concatenation function, $\sigma$ denotes a non-linear activation function, $\hat{\mathbb{S}}_{i}$ denotes the neighboring set of $S_i$ including self loop, and $\textbf{\emph{W}}^{l} \in\mathbb{R}^{p*q}$ is a scaling parameter matrix, $\textbf{\emph{a}}^{l+1} \in\mathbb{R}^{2q*1}$ is a scaling parameter vector for concatenated feature, and $\theta_{[S_i,S_j]}^{l+1}$ is the attention weight of edge $(S_i,S_j)$ in the $l+1$-th layer. Above attention process will be executed $k$ times in parallel to form the multi-head attention.

\subsection{Interaction Modules}
Inspired by the interaction mechanism of vision systems \cite{holtzman1984interactions,das1999topography,pecot2013multiple,cloutman2013interaction,medathati2016bio,milner2017two}, we design two kinds of interaction modules to enrich the features and enhance the discrimination ability of features.

\subsubsection{Self-interaction Module}
For the role of visual interactions in brain, a reasonable hypothesis is that internal self-interaction and feature information update exist between ventral stream and dorsal stream, and the self-interaction mechanism is more likely to be activated when processing biological visual information \cite{wei2020visual}.

\begin{figure}[htbp]
	\centering
	\includegraphics[width=0.7\linewidth]{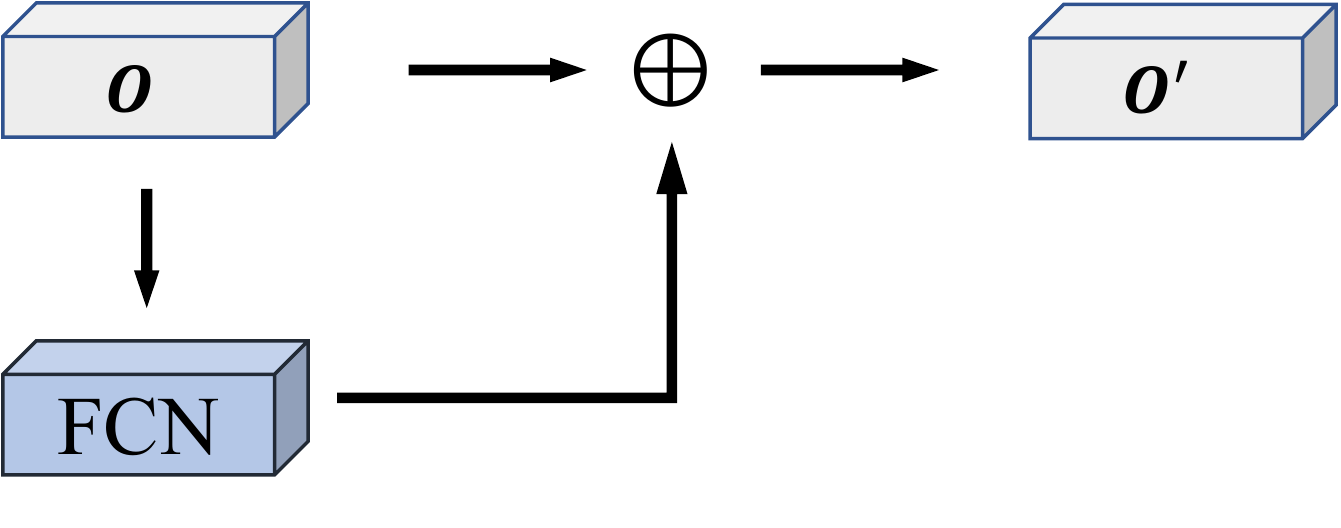}
	\caption{The illustration for a self-interaction module.}
	\label{fig2}
\end{figure}

We mimic self-interaction processing and propose a self-interaction module as shown in Fig. \ref{fig2}. For the input feature $\bm{O}$, it will return $\bm{O^{'}}$, which is the sum of itself with a recomputed vector from a fully connected network (FCN).

\subsubsection{Mutual-interaction Module}
Besides self-interaction processing, it has been proved that when dorsal stream and ventral stream process visual information, there are many mutual-interactions between these streams in each stage \cite{stettler2002lateral}. Similarly, inspired from the visual mutual interaction processing mechanism, we design a mutual-interaction module for our method to obtain a better integration of features from different streams. 

\begin{figure}[htbp]
	\centering
	\includegraphics[width=\linewidth]{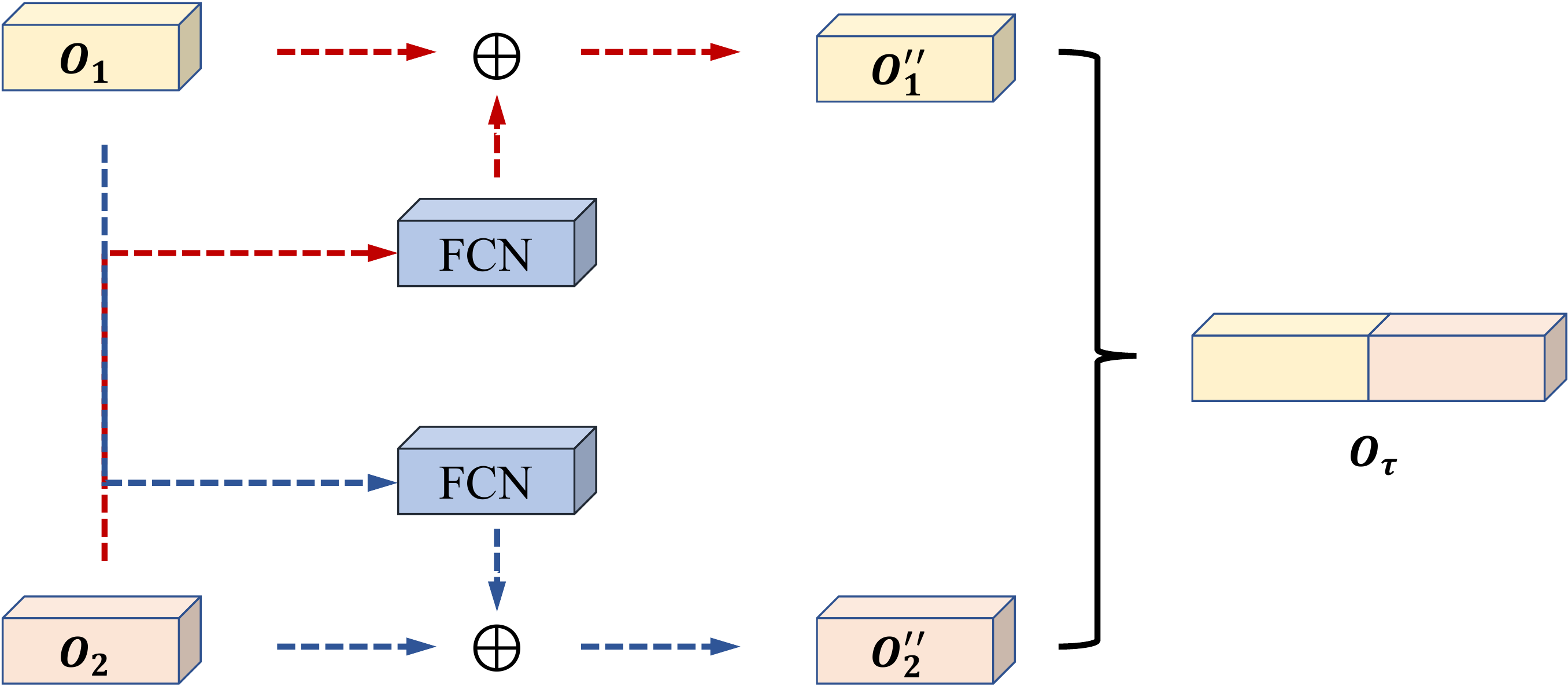}
	\caption{The illustration for a mutual-interaction module.}
	\label{fig3}
\end{figure}

As shown in Fig. \ref{fig3}, for features $\bm{O_1}$ and $\bm{O_2}$ from different streams, they will be input to a FCN independently to do scaling transformation and then return a sum with each other's original feature as follows,
\begin{equation}
\label{exn1}
\bm{O_i^{''}} = FCN(\bm{O_j}) + \bm{O_i}.
\end{equation}
Next the obtained $\bm{O_1^{''}}$ and $\bm{O_2^{''}}$ will be concatenated together to form the integration information $\bm{O_{\tau}}$. Through the mutual-interaction module, GNN-SEG can obtain the integrated features from different streams.

\subsection{Pixel-level Classification}
\begin{figure}[htbp]
	\centering
	\includegraphics[width=\linewidth]{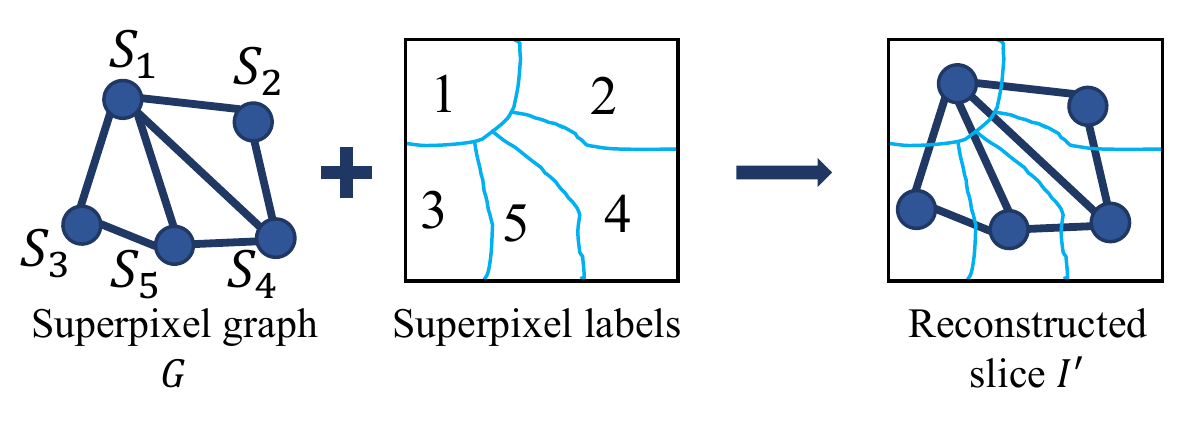}
	\caption{The illustration for transformation from a superpixel graph $G$ to a reconstructed slice $I^{'}$.}
	\label{fig4}
\end{figure}

After obtaining the integration information $\bm{O_{\tau}}$ for the superpixel graph $G$, we transform $G$ with its features $\bm{O_{\tau}}$ to a brain slice $I^{'}$ according to superpixel labels obtained by superpixel segmentation as Fig. \ref{fig4} shows. In $I^{'}$, each pixel in a superpixel shares the same feature of the superpixel. Then, $I^{'}$ will be combined into $I$ as an additional channel. Each pixel of the new $I$ will be classified by a pretrained pixel-level classifier. In this paper, we use a pretrained Attention Unet \cite{oktay2018attention} as the classifier. The parameters of the classifier are frozen.

\section{Experiments and results}
\label{expp}

\begin{table*}[htbp]
  \centering
  \scriptsize
  \caption{Comparison of performance of GNN-SEG and state-of-the-art methods on Brainweb. Best values are in bold. The italic values mean that they are better than the values obtained by CNN based methods.}
    \begin{tabular}{llllllllll}
    \hline
    \multicolumn{1}{c}{\multirow{2}[0]{*}{Methods}} &       & CSF   &       &       & GM    &       &       & WM    &  \\
          & Dice \% & TP \% & APD \%  & Dice \%  & TP \%    & APD \%  & Dice \%  & TP \%   & APD \%\\
    \hline
    \multicolumn{10}{c}{GNN-SEG variants}\\
    \hline
    GNN-SEG & \textbf{\emph{97.55$\pm$0.18}} & \textbf{\emph{97.79$\pm$0.95}} & \textbf{6.76$\pm$0.82} & \textbf{99.20$\pm$0.12} & 99.34$\pm$0.14 & \textbf{4.50$\pm$0.76} & \textbf{\emph{99.49$\pm$0.17}} & \textbf{\emph{99.50$\pm$0.35}} & \textbf{3.29$\pm$0.97} \\
    GNN-SEG (Fuzzy SLIC) & \textbf{\emph{97.46$\pm$0.22}} & \textbf{\emph{98.08$\pm$0.86}} & 8.00$\pm$1.56 & \textbf{\emph{99.15$\pm$0.15}} & 98.90$\pm$0.45 & \textbf{\emph{5.23$\pm$1.32}} & \textbf{\emph{99.48$\pm$0.09}} & \textbf{\emph{99.58$\pm$0.32}} & \textbf{\emph{3.51$\pm$0.90}} \\
    GNN-SEG (STC) & \textbf{\emph{97.55$\pm$0.12}} & \textbf{98.31$\pm$0.30} & 7.98$\pm$0.42 & \textbf{\emph{99.18$\pm$0.04}} & 99.10$\pm$0.22 & \textbf{\emph{5.19$\pm$0.50}} & \textbf{99.50$\pm$0.03} & \textbf{\emph{99.47$\pm$0.33}} & \textbf{\emph{3.42$\pm$0.31}} \\
    GNN-SEG (GCN) & \textbf{97.58$\pm$0.08} & \textbf{\emph{97.70$\pm$0.49}} & \textbf{\emph{6.80$\pm$0.45}} & \textbf{\emph{99.18$\pm$0.10}} & 98.78$\pm$0.31 & \textbf{\emph{4.83$\pm$0.52}} & \textbf{\emph{99.46$\pm$0.10}} & \textbf{99.78$\pm$0.07} & \textbf{\emph{3.60$\pm$0.94}} \\
    \hline
    \multicolumn{10}{c}{CNN based methods}\\
    \hline
    FCNN  & 87.91$\pm$0.26 & 87.69$\pm$1.15 & 34.07$\pm$0.55 & 93.44$\pm$0.06 & 93.93$\pm$0.74 & 33.68$\pm$0.77 & 95.27$\pm$0.06 & 95.16$\pm$0.46 & 29.70$\pm$0.57 \\
    Unet  & 96.81$\pm$0.44 & 96.63$\pm$0.77 & 9.18$\pm$0.91 & 98.49$\pm$0.45 & 99.26$\pm$0.57 & 9.06$\pm$2.35 & 98.67$\pm$0.63 & 97.69$\pm$1.56 & 8.56$\pm$3.69 \\
    SegNet & 93.10$\pm$1.41 & 95.35$\pm$1.90 & 23.64$\pm$8.51 & 96.14$\pm$0.69 & 94.87$\pm$1.39 & 22.39$\pm$5.38 & 97.26$\pm$0.49 & 98.16$\pm$0.64 & 20.94$\pm$7.51 \\
    Attention Unet & 97.20$\pm$0.25 & 96.97$\pm$1.34 & 7.76$\pm$0.88 & 99.06$\pm$0.04 & 99.14$\pm$0.47 & 5.26$\pm$0.12 & 99.29$\pm$0.03 & 99.05$\pm$0.71 & 4.37$\pm$0.50 \\
    Unet++ & 97.42$\pm$0.14 & 97.48$\pm$0.44 & 7.20$\pm$0.29 & 99.14$\pm$0.16 & \textbf{99.44$\pm$0.08} & 5.24$\pm$0.81 & 99.43$\pm$0.22 & 99.09$\pm$0.56 & 3.76$\pm$1.08 \\
    \hline
    \end{tabular}%
  \label{tab1}%
\end{table*}%

\begin{table*}[htbp]
  \centering
    \scriptsize
  \caption{Comparison of performance of GNN-SEG and state-of-the-art methods on MRBrainS. Best values are in bold. The italic values mean that they are better than the values obtained by CNN based methods.}
    \begin{tabular}{llllllllll}
    \hline
    \multicolumn{1}{c}{\multirow{2}[0]{*}{Methods}} &       & CSF   &       &       & GM    &       &       & WM    &  \\
          & Dice \% & TP \% & APD \%  & Dice \%  & TP \%    & APD \%  & Dice \%  & TP \%   & APD \%\\
    \hline
    \multicolumn{10}{c}{GNN-SEG variants}\\
    \hline
    GNN-SEG & \textbf{\emph{91.85$\pm$0.65}} & \textbf{\emph{92.63$\pm$1.80}} & 34.94$\pm$0.65 & \textbf{\emph{88.59$\pm$1.23}} & \textbf{\emph{88.12$\pm$1.63}} & \textbf{\emph{55.16$\pm$5.43}} & \textbf{\emph{86.80$\pm$0.74}} & 88.19$\pm$0.17 & \textbf{\emph{52.79$\pm$3.82}} \\      
    GNN-SEG (Fuzzy SLIC) & \textbf{\emph{92.24$\pm$0.27}} & \textbf{94.19$\pm$1.26} & 35.65$\pm$2.26 & \textbf{\emph{88.21$\pm$0.75}} & \textbf{\emph{88.41$\pm$1.23}} & \textbf{51.70$\pm$1.19} & \textbf{\emph{86.98$\pm$0.49}} & 88.14$\pm$0.53 & \textbf{\emph{48.94$\pm$2.51}} \\
    GNN-SEG (STC) & \textbf{92.36$\pm$0.29} & \textbf{\emph{92.86$\pm$0.80}} & \textbf{32.95$\pm$1.23} & \textbf{89.34$\pm$0.40} & \textbf{89.67$\pm$2.00} & \textbf{\emph{53.17$\pm$2.00}} & \textbf{87.59$\pm$0.09} & 87.81$\pm$0.19 & \textbf{46.97$\pm$1.48} \\
    GNN-SEG (GCN) & \textbf{\emph{91.80$\pm$0.16}} & \textbf{\emph{92.19$\pm$1.77}} & 36.82$\pm$3.36 & \textbf{\emph{88.22$\pm$0.54}} & \textbf{\emph{88.18$\pm$3.90}} & \textbf{\emph{55.82$\pm$3.61}} & \textbf{\emph{86.73$\pm$0.65}} & 87.58$\pm$2.68 & \textbf{\emph{52.73$\pm$2.72}} \\
    \hline
    \multicolumn{10}{c}{CNN based methods}\\
    \hline
    FCNN  & 86.43$\pm$0.54 & 86.07$\pm$1.63 & 59.39$\pm$4.11 & 82.70$\pm$0.38 & 81.44$\pm$1.36 & 86.06$\pm$2.12 & 80.36$\pm$0.36 & 82.15$\pm$1.27 & 71.97$\pm$3.13 \\
    Unet  & 91.77$\pm$0.50 & 90.73$\pm$2.43 & 34.53$\pm$0.94 & 88.02$\pm$0.47 & 86.55$\pm$3.47 & 56.32$\pm$3.31 & 86.72$\pm$0.45 & \textbf{89.88$\pm$1.40} & 53.92$\pm$1.89 \\
    SegNet & 90.09$\pm$0.42 & 89.96$\pm$1.82 & 43.29$\pm$0.25 & 85.30$\pm$0.57 & 87.93$\pm$1.57 & 86.60$\pm$11.72 & 82.61$\pm$1.06 & 81.61$\pm$1.54 & 62.78$\pm$2.70 \\
    Attention Unet & 91.69$\pm$0.46 & 92.11$\pm$1.15 & 38.34$\pm$2.64 & 87.13$\pm$0.96 & 87.97$\pm$1.00 & 61.67$\pm$6.89 & 85.27$\pm$1.35 & 84.48$\pm$1.85 & 54.65$\pm$5.81 \\
    Unet++ & 91.44$\pm$0.43 & 91.05$\pm$3.84 & 37.11$\pm$6.13 & 87.88$\pm$0.88 & 87.34$\pm$3.38 & 58.94$\pm$1.52 & 86.27$\pm$1.09 & 87.73$\pm$3.07 & 53.53$\pm$2.43 \\
    \hline
    \end{tabular}%
  \label{tab2}%
\end{table*}%

\begin{table*}[htbp]
  \centering
    \scriptsize
  \caption{Comparison of performance of GNN-SEG and state-of-the-art methods on IBSR. Best values are in bold. The italic values mean that they are better than the values obtained by CNN based methods.}
    \begin{tabular}{llllllllll}
    \hline 
    \multicolumn{1}{c}{\multirow{2}[0]{*}{Methods}} &       & CSF   &       &       & GM    &       &       & WM    &  \\
          & Dice \% & TP \% & APD \%  & Dice \%  & TP \%    & APD \%  & Dice \%  & TP \%   & APD \%\\
    \hline
    \multicolumn{10}{c}{GNN-SEG variants}\\
    \hline     
    GNN-SEG & \textbf{\emph{81.34$\pm$0.07}} & \textbf{\emph{79.77$\pm$0.36}} & \textbf{40.86$\pm$0.39} & \textbf{\emph{96.20$\pm$0.07}} & 96.72$\pm$0.51 & \textbf{\emph{35.28$\pm$0.29}} & \textbf{\emph{95.95$\pm$0.10}} & \textbf{\emph{95.73$\pm$0.48}} & \textbf{\emph{32.98$\pm$1.44}} \\          
    GNN-SEG (Fuzzy SLIC) & \textbf{\emph{81.43$\pm$0.17}} & \textbf{\emph{80.01$\pm$0.50}} & 41.83$\pm$1.29 & \textbf{\emph{96.22$\pm$0.07}} & 96.78$\pm$0.39 & 36.02$\pm$1.40 & \textbf{\emph{95.93$\pm$0.11}} & \textbf{\emph{95.44$\pm$0.65}} & \textbf{\emph{33.85$\pm$1.28}} \\
    GNN-SEG (STC) & \textbf{81.53$\pm$0.12} & \textbf{\emph{80.18$\pm$0.26}} & 41.94$\pm$1.21 & \textbf{\emph{96.24$\pm$0.04}} & \textbf{97.07$\pm$0.25} & 36.15$\pm$0.55 & \textbf{\emph{95.96$\pm$0.08}} & \textbf{\emph{95.39$\pm$0.56}} & \textbf{\emph{33.62$\pm$0.78}} \\
    GNN-SEG (GCN) & \textbf{\emph{81.45$\pm$0.14}} & \textbf{80.74$\pm$0.91} & 41.49$\pm$1.38 & \textbf{96.36$\pm$0.03} & 96.60$\pm$0.20 & \textbf{33.47$\pm$1.08} & \textbf{96.13$\pm$0.05} & \textbf{95.98$\pm$0.15} & \textbf{31.45$\pm$0.30} \\
    \hline
    \multicolumn{10}{c}{CNN based methods}\\
    \hline
    FCNN  & 70.53$\pm$0.11 & 69.10$\pm$0.36 & 72.88$\pm$3.46 & 93.25$\pm$0.13 & 93.71$\pm$0.49 & 68.32$\pm$2.94 & 92.66$\pm$0.15 & 92.54$\pm$0.59 & 63.51$\pm$2.68 \\
    Unet  & 80.11$\pm$0.05 & 77.81$\pm$1.52 & 40.95$\pm$1.65 & 96.01$\pm$0.13 & 96.65$\pm$0.74 & 37.12$\pm$1.48 & 95.67$\pm$0.31 & 95.20$\pm$1.40 & 34.87$\pm$1.86 \\
    SegNet & 76.06$\pm$0.84 & 73.47$\pm$2.41 & 48.49$\pm$1.94 & 95.10$\pm$0.07 & 95.75$\pm$0.21 & 44.20$\pm$0.54 & 94.54$\pm$0.11 & 93.95$\pm$0.67 & 42.94$\pm$1.06 \\
    Attention Unet & 80.09$\pm$0.42 & 78.66$\pm$1.15 & 42.91$\pm$0.52 & 96.09$\pm$0.04 & 97.05$\pm$0.14 & 35.74$\pm$1.07 & 95.69$\pm$0.03 & 94.64$\pm$0.07 & 33.91$\pm$0.53 \\
    Unet++ & 80.42$\pm$0.03 & 78.17$\pm$0.54 & 40.91$\pm$0.78 & 96.08$\pm$0.10 & 96.69$\pm$0.37 & 35.46$\pm$0.62 & 95.64$\pm$0.28 & 95.00$\pm$0.98 & 34.28$\pm$0.89 \\
    \hline 
    \end{tabular}%
  \label{tab3}%
\end{table*}%

\begin{table*}[htbp]
  \centering
    \scriptsize
  \caption{Comparison of performance of GNN-SEG and state-of-the-art methods on iSeg-2019. Best values are in bold. The italic values mean that they are better than the values obtained by CNN based methods.}
    \begin{tabular}{llllllllll}
    \hline
    \multicolumn{1}{c}{\multirow{2}[0]{*}{Methods}} &       & CSF   &       &       & GM    &       &       & WM    &  \\
          & Dice \% & TP \% & APD \%  & Dice \%  & TP \%    & APD \%  & Dice \%  & TP \%   & APD \%\\
    \hline
    \multicolumn{10}{c}{GNN-SEG variants}\\
    \hline
    GNN-SEG & \textbf{\emph{92.97$\pm$0.14}} & 92.64$\pm$1.52 & 23.34$\pm$1.82 & \textbf{91.63$\pm$0.07} & \textbf{\emph{92.00$\pm$0.37}} & \textbf{\emph{43.59$\pm$0.46}} & \textbf{\emph{90.82$\pm$0.03}} & \textbf{90.90$\pm$0.27} & \textbf{\emph{48.11$\pm$0.22}} \\      
    GNN-SEG (Fuzzy SLIC) & \textbf{\emph{92.98$\pm$0.04}} & 92.34$\pm$0.53 & 22.52$\pm$1.30 & \textbf{\emph{91.48$\pm$0.07}} & \textbf{92.02$\pm$0.69} & \textbf{\emph{44.00$\pm$0.19}} & \textbf{\emph{90.69$\pm$0.14}} & \textbf{\emph{90.43$\pm$1.02}} & \textbf{\emph{48.76$\pm$0.47}} \\
    GNN-SEG (STC) & \textbf{93.06$\pm$0.01} & \textbf{93.52$\pm$0.37} & 23.40$\pm$1.48 & \textbf{\emph{91.62$\pm$0.06}} & \textbf{\emph{91.84$\pm$0.49}} & \textbf{43.59$\pm$0.39} & \textbf{90.87$\pm$0.01} & \textbf{\emph{90.63$\pm$0.10}} & \textbf{47.98$\pm$0.34} \\
    GNN-SEG (GCN) & \textbf{\emph{93.02$\pm$0.11}} & 93.14$\pm$0.55 & 23.67$\pm$1.14 & \textbf{\emph{91.51$\pm$0.02}} & \textbf{\emph{91.99$\pm$0.25}} & \textbf{\emph{44.16$\pm$0.15}} & \textbf{\emph{90.72$\pm$0.07}} & \textbf{\emph{90.40$\pm$0.28}} & \textbf{\emph{48.33$\pm$0.63}} \\
    \hline
    \multicolumn{10}{c}{CNN based methods}\\
    \hline
    FCNN  & 89.42$\pm$0.07 & 88.30$\pm$0.18 & 31.36$\pm$0.14 & 88.41$\pm$0.05 & 89.10$\pm$0.40 & 56.55$\pm$0.17 & 87.19$\pm$0.11 & 86.84$\pm$0.66 & 64.23$\pm$0.41 \\
    Unet  & 92.73$\pm$0.15 & 91.19$\pm$0.46 & \textbf{21.96$\pm$0.55} & 91.17$\pm$0.21 & 91.36$\pm$0.44 & 44.94$\pm$1.01 & 90.40$\pm$0.16 & 90.35$\pm$0.24 & 51.15$\pm$1.40 \\
    SegNet & 91.44$\pm$0.28 & 91.06$\pm$0.65 & 26.55$\pm$1.53 & 88.94$\pm$0.37 & 89.81$\pm$0.83 & 53.29$\pm$2.13 & 87.39$\pm$0.80 & 86.40$\pm$2.63 & 62.14$\pm$1.92 \\
    Attention Unet & 92.69$\pm$0.19 & 93.45$\pm$1.35 & 25.55$\pm$2.26 & 91.25$\pm$0.13 & 91.58$\pm$0.27 & 44.79$\pm$0.15 & 90.47$\pm$0.11 & 89.76$\pm$0.36 & 48.89$\pm$0.47 \\
    Unet++ & 92.94$\pm$0.09 & 92.50$\pm$0.34 & 22.93$\pm$0.28 & 91.46$\pm$0.11 & 91.71$\pm$0.15 & 44.24$\pm$0.71 & 90.66$\pm$0.08 & 90.28$\pm$0.58 & 49.04$\pm$1.38 \\
    \hline
    \end{tabular}%
  \label{tab4}%
\end{table*}%

\begin{figure*}[htbp]
	\centering
	\includegraphics[width=0.8\linewidth]{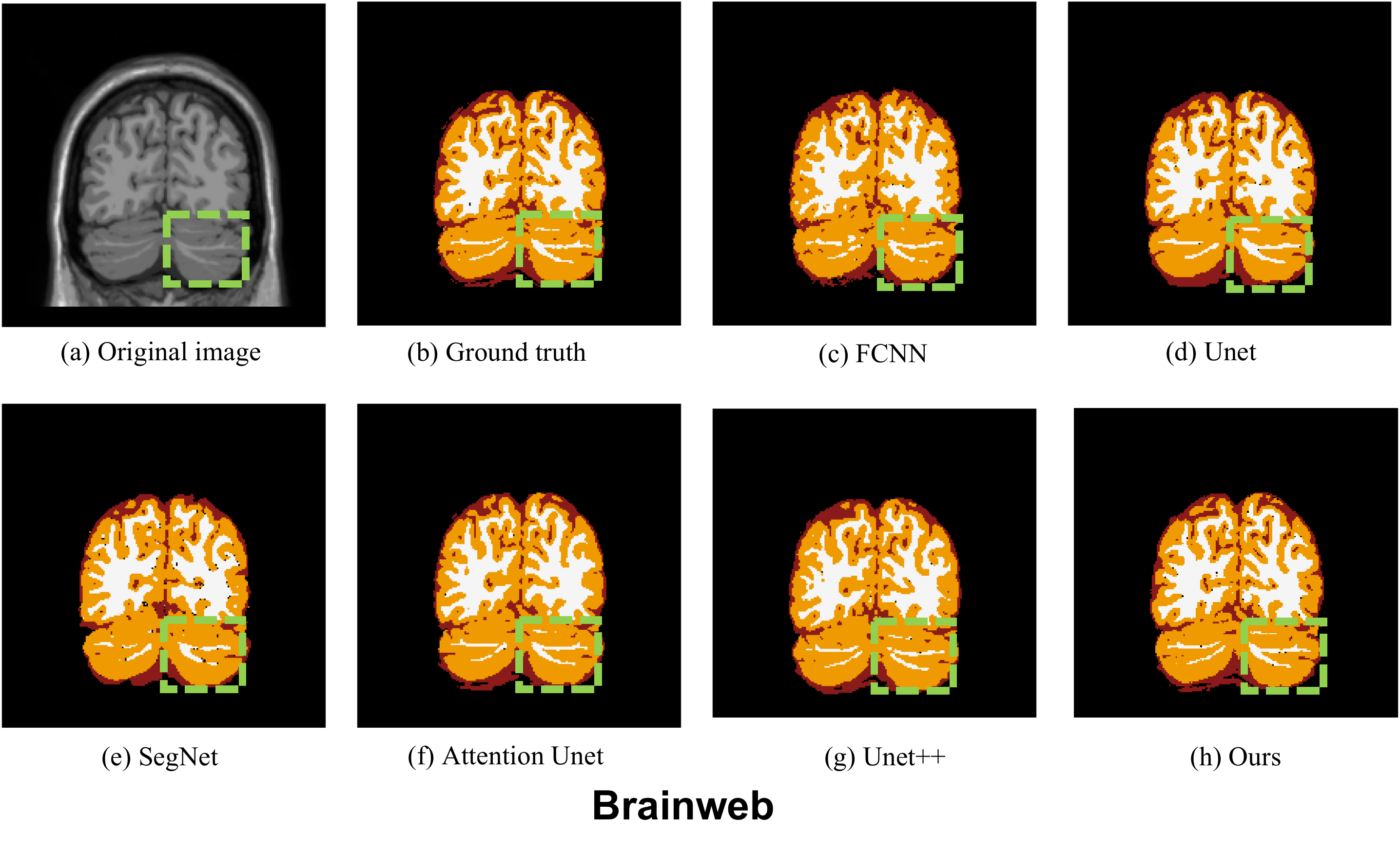}
	\caption{Visual comparison of segmentation results of GNN-SEG and state-of-the-art methods on Brainweb.}
	\label{fig51}
\end{figure*}

\begin{figure*}[htbp]
	\centering
	\includegraphics[width=0.8\linewidth]{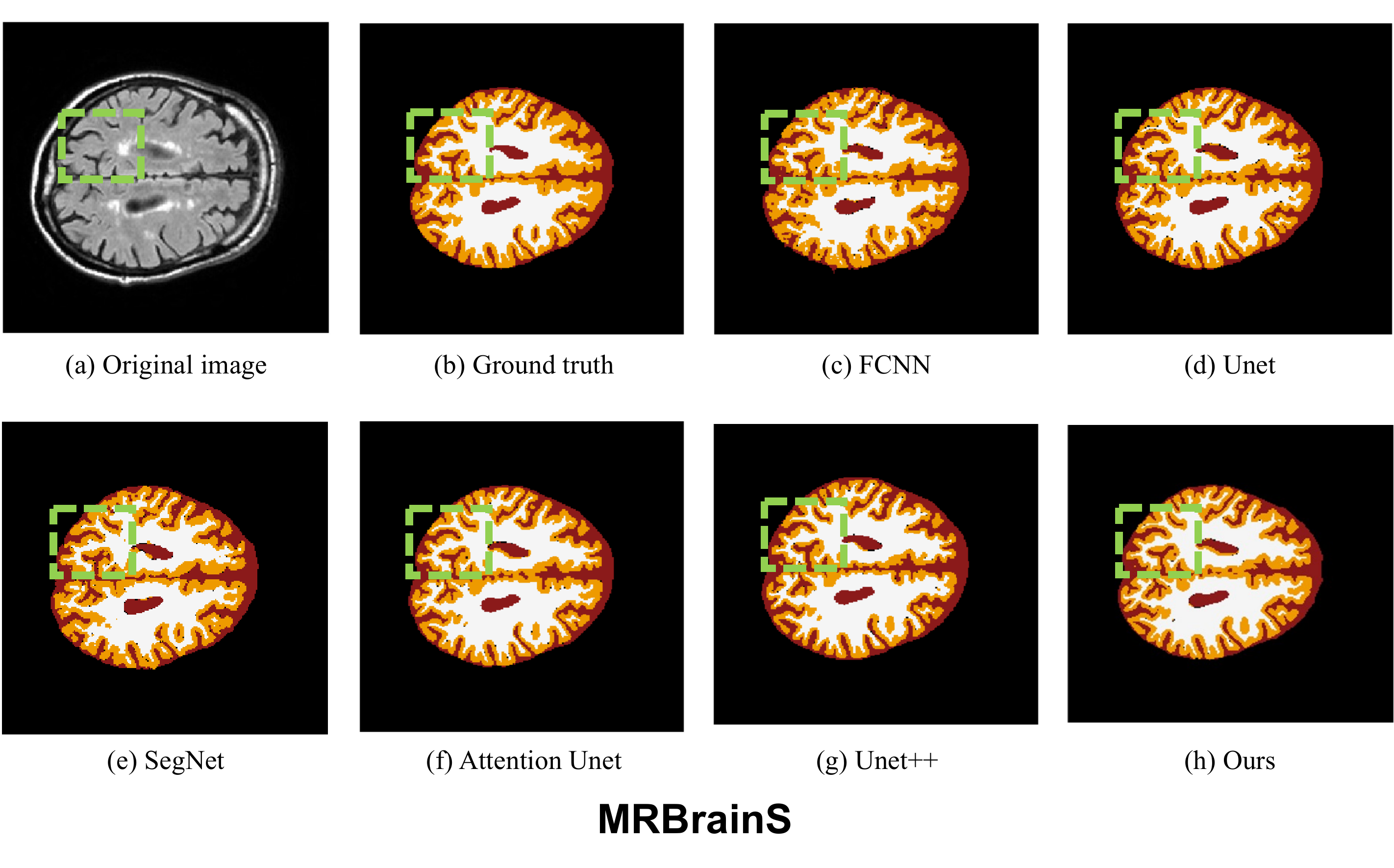}
	\caption{Visual comparison of segmentation results of GNN-SEG and state-of-the-art methods on MRBrainS.}
	\label{fig52}
\end{figure*}

\begin{figure*}[htbp]
	\centering
	\includegraphics[width=0.8\linewidth]{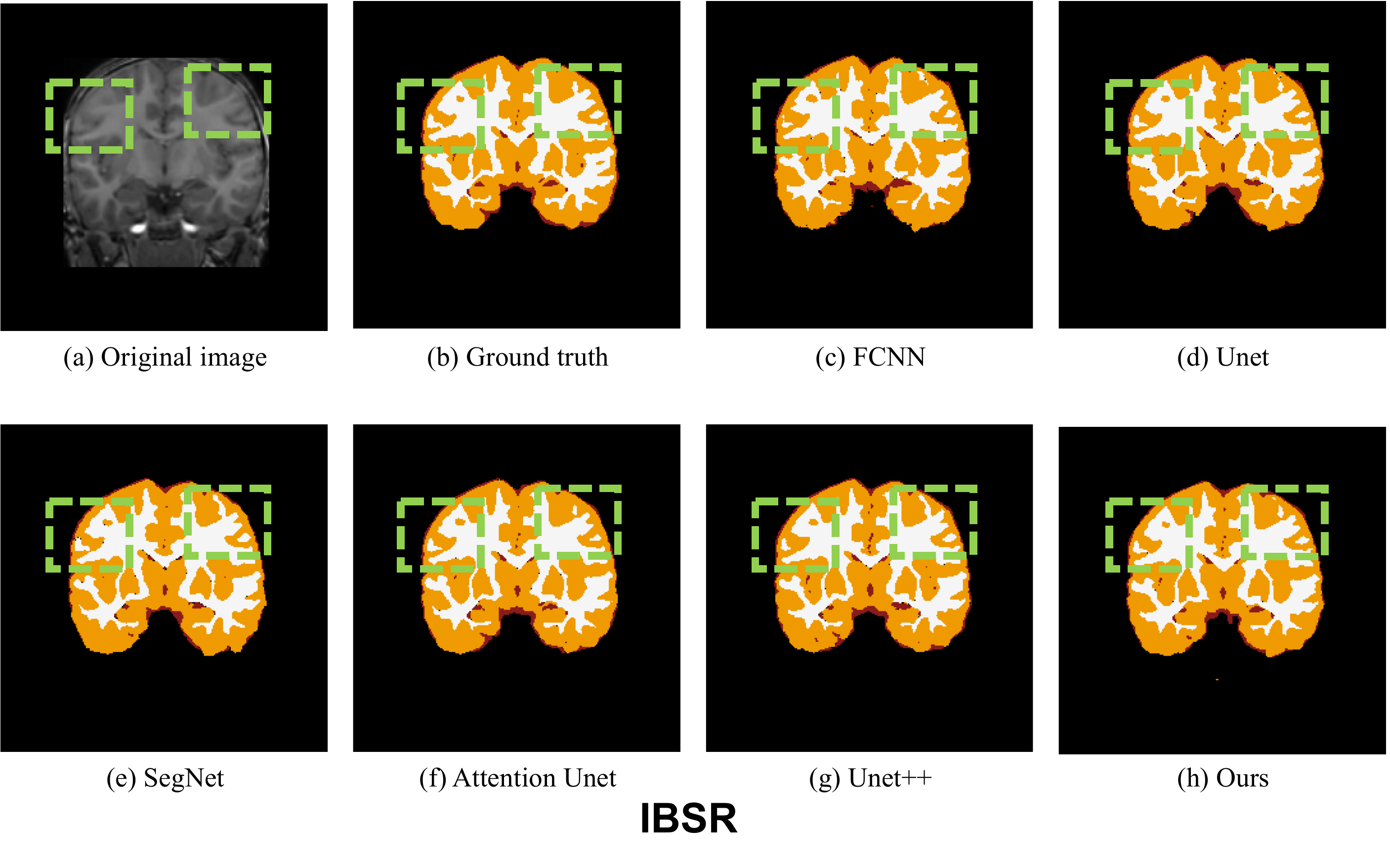}
	\caption{Visual comparison of segmentation results of GNN-SEG and state-of-the-art methods on IBSR.}
	\label{fig53}
\end{figure*}

\begin{figure*}[htbp]
	\centering
	\includegraphics[width=0.8\linewidth]{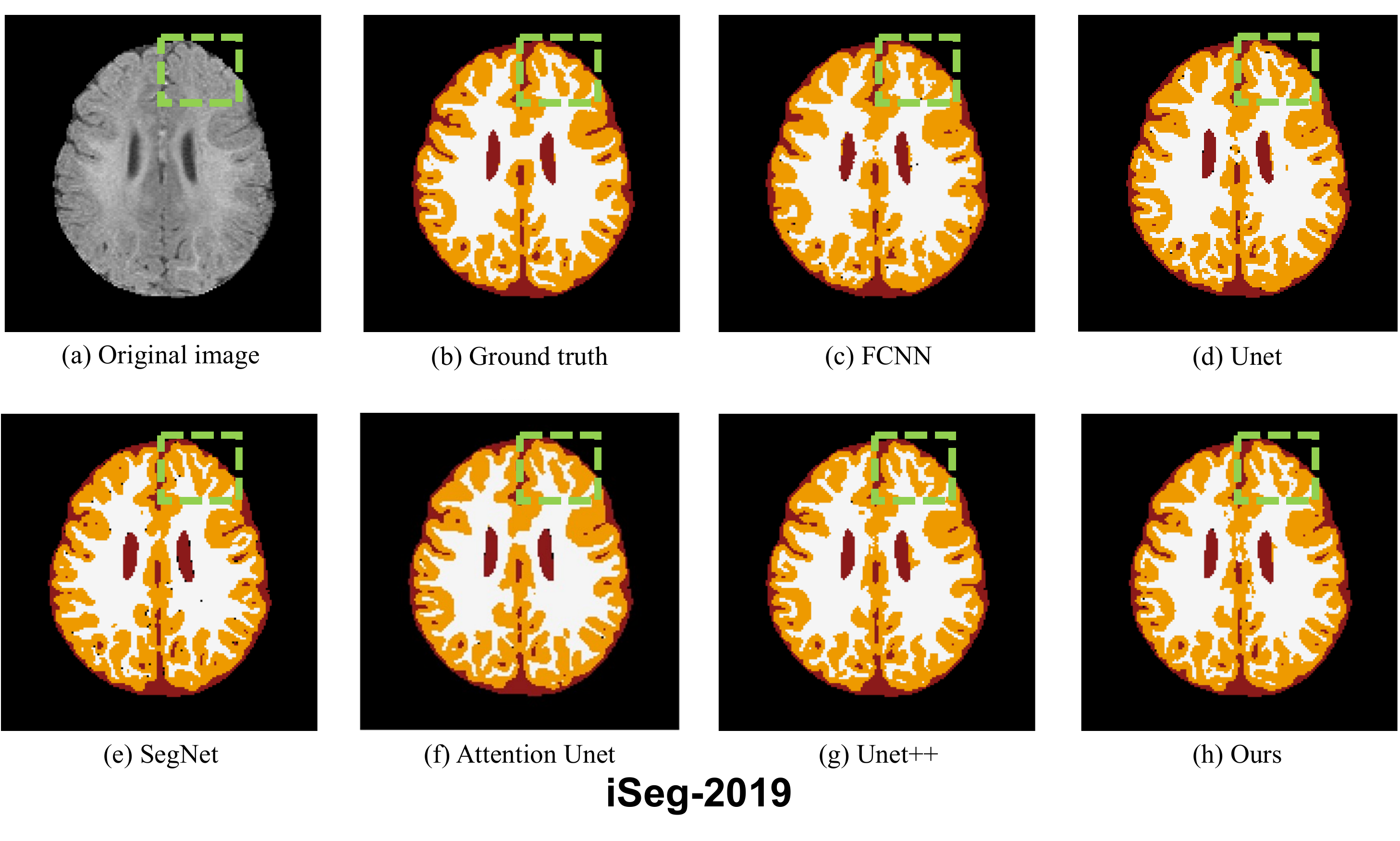}
	\caption{Visual comparison of segmentation results of GNN-SEG and state-of-the-art methods on iSeg-2019.}
	\label{fig54}
\end{figure*}

\begin{table}[htbp]
  \centering
  \caption{Number of trainable parameters of GNN-SEG and state-of-the-art methods on MRBrainS.}
    \begin{tabular}{llllll}
    \hline
    GNN-SEG & FCNN  & Unet  & SegNet & Attention Unet & Unet++ \\
    \hline
    0.1M & 20.5M & 31.0M & 25.0M & 34.9M & 9.3M \\
    \hline
    \end{tabular}%
  \label{tab5}%
\end{table}%

\subsection{Datasets}
For the experiments, we evaluate our method using four brain MRI datasets, which are acquired from Brainweb dataset \cite{cocosco1997brainweb}, MRBrainS dataset \cite{mendrik2015mrbrains}, iSeg-2019 dataset \cite{sun2021multi}, and IBSR dataset \cite{valverde2015comparison}. Brainweb is a synthetic brain MRI dataset. We selected 370 slices. Among them, 240 slices were used for training and 130 slices were used for testing. Each slice has three modalities: T1w (T1 weighted), T2w (T2 weighted), and PD (proton density weighted). MRBrainS dataset was downloaded from the grand challenge on brain MRI segmentation at MICCAI 2018. We selected 135 slices. Among them, 95 slices were used for training and 40 slices were used for testing. Each slice has three modalities: T1w, T1-IR (T1 weighted inversion recovery), and T2-FLAIR (T2 weighted fluid attenuated inversion recovery). iSeg-2019 dataset was downloaded from the MICCAI Grand Challenge on 6-month infant brain MRI segmentation from multiple sites. We selected 500 slices. Among them, 350 slices were used for training and 150 slices were used for testing. Each slice has two modalities: T1w and T2w. IBSR dataset consists of 18 real brain MRIs derived from healthy subjects. We selected 1440 slices. Among them, 1000 slices were used for training and 440 slices were used for testing. Each slice has only one modality: T1w. Although the segmentation benchmarks in above four datasets include different brain tissues, for all of them, we classify them into cerebrospinal fluid (CSF), gray matter (GM), and white matter (WM) as target segmentation objects for comparison.

\subsection{Experiment Settings}
To validate our method, GNN-SEG was compared with several state-of-the-art CNN based deep learning segmentation methods: FCNN \cite{long2015fully}, Unet \cite{ronneberger2015u}, SegNet \cite{badrinarayanan2017segnet}, Attention Unet \cite{oktay2018attention}, and Unet++ \cite{zhou2018unet++}. The optimizer for GNN-SEG was Adam with a learning rate 0.001. Besides, all comparison methods used in this paper adopted their default settings. To clarify the effect of different superpixel methods on the performance of GNN-SEG, we selected fuzzy simple linear iterative clustering (Fuzzy SLIC) \cite{fslic} as the comparison. The corresponding GNN-SEG variant is called GNN-SEG (Fuzzy SLIC). What's more, to clarify the effect of different GNNs on the performance of GNN-SEG, two variants of GNN-SEG called GNN-SEG (STC) and GNN-SEG (GCN), which are based on star topology convolution (STC) \cite{STC} and graph convolutional network (GCN) \cite{GCN} respectively, were introduced for comparison. The settings of three variants were the same as GNN-SEG. All experiments were run on a personal computer with Windows OS 10 Home, AMD Ryzen 9 5900HX 3.30 GHz 8-Core CPU, 32 GB RAM, CUDA version 11.4.141, torch version 1.10.0, and 1 NVIDIA GeForce RTX 3080 Laptop (16 GB) GPU.

\subsection{Evaluation Metrics}
Three metrics are selected to evaluate the performance of each method: Sensitivity (TP), Dice, and average perpendicular distance (APD). TP, Dice, and APD can be calculated as follows,
\begin{equation}
 {\rm{TP}} = \frac{\left | P_1\bigcap T_1 \right |}{\left |T_1  \right |},
\label{eTP}
\end{equation}
\begin{equation}
 {\rm{Dice}} = \frac{\left | P_1\bigcap T_1 \right |}{(\left |T_1  \right | + \left |P_1  \right |)/2},
\label{eDice}
\end{equation}
\begin{equation}
 {\rm{APD}} = \frac{\sum_{i=1}^{\left | P_d \right |} D(P_{d,[i]}, T_d)}{\left | P_d \right |},
\label{eAPD}
\end{equation}
where, $T_1$ is the number of pixels of the object tissue within the benchmark, and $P_1$ is the number of pixels of the predicted object tissue in the slice, $T_d$ and $P_d$ are boundaries of the benchmark and the segmentation result, respectively, $\left | P_d \right |$ is the number of pixels of $P_d$, and $D(P_{d,[i]}, T_d)$ will return the minimum Euclidean distance between $T_d$ and the $i$th pixel in $P_d$. TP and Dice mainly measure the overlapping degree between the predicted results and the actual objects, and APD mainly calculates the distance between two boundaries. Hence, higher scores of TP and Dice and lower score of APD mean better performance.

\subsection{Performance Analysis}
Table \ref{tab1} shows the numerical comparison of segmentation results obtained by all methods in terms of Dice, TP, and APD on Brainweb. It can be seen that GNN-SEG and its variants outperformed CNN based methods in the segmentation of all three tissues. Unet, Unet++, and Attention Unet achieved close result compared to GNN-SEG and its variants. Because Brainweb is a synthetic brain MRI dataset, the difference between our methods and conventional CNN based methods is not significant. To further validate the performance of the proposed GNN-SEG, we applied it on three real brain MRI datasets: MRBrainS, IBSR, and iSeg-2019. The numerical comparison of segmentation results of GNN-SEG and state-of-the-art methods are as Tables \ref{tab2}-\ref{tab4} show. It can be seen that the overall performance of GNN-SEG and its variants is still better than CNN based methods on these three real brain MRI datasets. What's more, GNN-SEG and its variants outperformed CNN based methods significantly on MRBrainS and IBSR. The difference between the performance of GNN-SEG and two variants using different GNNs is not significant. Similarly, the difference between the performance of GNN-SEG and the variant using different superpixel method is also not significant. Figs. \ref{fig51}-\ref{fig54} show the visual comparison of GNN-SEG and state-of-the-art methods on Brainweb, MRBrainS, IBSR, and iSeg-2019. It can be seen that benefit from the rich structural information learned by structural relation learning model of GNN-SEG, the segmentation of tissue branches obtained by GNN-SEG is more accurate than state-of-the-art CNN based methods. Table \ref{tab5} shows the number of trainable parameters of GNN-SEG and state-of-the-art methods. It can be seen that GNN-SEG has less trainable parameters than state-of-the-art methods.

\section{Conclusion}
In this paper, we propose a novel brain tissue segmentation method which takes superpixels as basic processing units and uses graph neural networks to learn the relation between different superpixels. In addition, we mimic the interaction mechanism of biological vision systems and design two kinds of interaction modules for feature enhancement and integration. The experimental results show that our method has better segmentation performance than state-of-the-art CNN based methods on four brain MRI datasets. In the future, we plan to validate our method on other biomedical image segmentation tasks.

\section{Acknowledgements}
This work is supported by Hong Kong Innovation and Technology Commission (InnoHK Project CIMDA), Hong Kong Research Grants Council (Project 11204821), and City University of Hong Kong (Project 9610034).

\bibliographystyle{IEEEbib}
\bibliography{mybibfile}

\end{document}